\documentclass[aps,prl,reprint, groupedaddress]{revtex4-1}

\usepackage{graphicx}
\usepackage{dcolumn}
\usepackage{bm}

\begin{document}


\title{Molecular dynamics (MD) calculation of the zeta potential of neutral surfaces}


\author{Hongyi Liu}
\email[]{hl373@cornell.edu}
\affiliation{Department of Earth and Atmospheric Science, Cornell University, Ithaca, NY 14853}
\affiliation{The KAUST-Cornell Center for Energy and Sustainability, Cornell University, Ithaca, NY 14853}

\author{Lawrence M. Cathles, III}
\email[]{lmc19@cornell.edu}
\affiliation{Department of Earth and Atmospheric Science, Cornell University, Ithaca, NY 14853}
\affiliation{The KAUST-Cornell Center for Energy and Sustainability, Cornell University, Ithaca, NY 14853}


\date{\today}

\begin{abstract}
Molecular dynamics (MD) simulations of the zeta potential are so poor that it has become common to term their predictions 'apparent'. Here we demonstrate how zeta potentials that agree with measured values can be calculated by: (1) integrating the net average charge in surface-parallel layers from the midpoint of the fluid layer (where the electrostatic potential is zero) to and then into two solid caps, (2) determining the position of slipping plane with separate Couette flow models, and (3) calculating the charge distribution and electrostatic potential under static conditions. The solids we model are charge neutral surfaces composed of atoms with zero charge or charge balanced monovalent or divalent ions. The zeta potentials calculated are within a few millivolts of measured values, and the measured values fall within the simulation error bars. Insights provided by the improved MD simulations into the complex phenomena that affect surface charge and zeta potential are discussed.
\end{abstract}

\pacs{}

\maketitle

The zeta potential, the electrostatic potential at the plane that separates the mobile and stationary parts of a fluid phase in contact a solid \cite{a03, a04, a05, a06}, is a fundamental parameter that is clearly defined, easily measured, and important. For example, it controls how particles aggregate \cite{a05, a06}. The zeta potential is not, however, easily modeled. MD implementations of the classical Helmholtz and Smoluchowski (H-S) theory that is successful at the laboratory scale \cite{a01} have failed to predict the zeta potential measurements \cite{a03, a04, a07, a08, ajj}. For example, Huang \textit{et al.} \cite{a03} predicted -38 mV for the zeta potential of a zero-charge liposomes bio-surface in 1M NaI solution whereas measurement in a similar KI solution indicates $\sim$$-$9 mV \cite{a09}. Zeta potentials calculated by MD methods are commonly referred to as apparent (as opposed to real) \cite{a04}. Other theoretical techniques such as atomic scale \cite{a10, a11} and Monte Carlo simulations \cite{a12,a13} have also encountered prediction difficulties.

The current method for calculating zeta potentials (e.g., Huang \textit{et al.} \cite{a03}) uses MD methods to calculate the viscosity and rate of fluid flow past the solid interface, and then uses the H-S equation to compute the zeta potential. Two problems are obvious: First, the cumulative thickness of the fluid layers with non-zero net ion charge is typically $\sim$20\% of the separation between the MD solid surfaces which is too large for the assumptions grounding the H-S equation. Second, MD fluid velocities are 100’s of times greater than those which occur in the laboratory or in nature \cite{a03, a18}, and their high near-surface shear changes the solution viscosity and charge distribution significantly.

In this paper we calculate the zeta potentials for the simplest possible situation of charge-neutral surfaces, and show how it is possible to calculate zeta potentials which agree with experimental measurements.

Our approach follows the pioneering methods of Spohr \cite{a14, a15} who obtained the mean electrostatic potential in the fluid phase by 1D integration of the charge distribution along lines perpendicular to a charge-neutral mercury surface. He predicted unrealistically high electrostatic potentials of 430 mV at 20 \AA\ from the solid surface, and non-physical charge density variations nearer to the surface. Following Spohr’s methods Lorenz \textit{et al.} \cite{a18} and Spagnoli \textit{et al.} \cite{a19} also obtained near-surface electrostatic well outside the observed range of $-$100 to $+$100 mV \cite{a16, a17}, and also found strong charge oscillations near the solid surface. Huang \textit{et al.} \cite{a03} suggested integrating charge from the middle of the solution layer (where it is assumed that the electrostatic potential is zero). But Huang \textit{et al.} calculated the apparent zeta potential using the H-S equation.

Here we apply the methods suggested (but not used) by Huang \textit{et al.} with only slight (but important) modifications to calculate the zeta potentials adjacent to charge-neutral surfaces. The non-zero zeta potentials observed  \cite{a09, a20, a21, a22, a23} and calculated \cite{a03, a04, a07, a08, a18, a24} for charge-neutral surfaces are thought to result from ions binding to the surface \cite{a09, a22}, reduction of ion mobility in the hydration layer \cite{a24}, or a mobility difference between positive and negative ions \cite{a03}. Our modeling indicates the first is most likely.

\textit{Modeling Methods.} Solids composed of numerical atoms with no charge, charge-balanced monovalent atoms, and charge-balanced divalent atoms are considered. Three unit cell layers of 648 atoms are arranged in a perfect FCC lattice with the (1 0 0) face exposed to the numerical aqueous saline solution. The aqueous solution lies between two solid caps, and the mid-plane of the solution layer is assumed to have zero electrostatic potential. We define the solid surface by the centers of its first layer of solid atoms. The force field of K$^{+}$ is adopted from the work of Koneshan \textit{et al.} \cite{a25}. The hydrophobic surface is from Huang \textit{et al.} \cite{a03}. The electrolyte salinity is approximately 1M (40 ion pairs in 2160 SPC/E model water molecules). NaCl, NaI, KCl and KI brines considered. Valid simulations require the fluid layer to be at least 4.1 nm thick \cite{a24} (ours is $\sim$5 nm), and the non-bound fluid layer must be at least 30\% of the total \cite{a26, a27}. Our simulations satisfy these criteria.

We employ a 2-D periodic boundary condition in the horizontal x and y directions. The long-range electrostatic force is calculated using the PPPM method with slab option 3.0 \cite{a28}. The system is first equilibrated to P$_{z}$ = 10 atm and T=298 K by running the simulator for 5 ns under the ensemble of NP$_{z}$T. An additional 15 ns of simulation under static conditions provides the results for the static model (SM). All simulations were performed with the LAMMPS MD simulation package \cite{a29} and, except for the integration procedure, use the same methods as described by Huang \textit{et al.} \cite{a03}.

Equation (1) shows how the net charge density, $\sigma$, is determined by averaging the MD ion distributions in a thin layer of thickness $dz'$ a distance $z_{m}-z’$ from the the mid-plane ($z=z_{m}$) and integrating to obtain the electrostatic potential in the fluid:
\begin{equation}
\Phi(z)=-\frac{1}{2\cdot\varepsilon_{w}\cdot\varepsilon_{0}}\intop_{z}^{z_{m}}dz'(z_{m}-z')\cdot\sigma(z')
\end{equation}
where the relative dielectric constant of water at room temperature is $\varepsilon_{w}$, the absolute dielectric constant $\varepsilon_{0}$, and $\sigma(z')$ is the charge density in Coulombs per unit area at the position $z'$.

Flowing conditions are induced by several methods. For the Couette flow  (CF) models, a simple linear shear is imposed on the fluid. The velocities of the top solid surfaces is $\pm$5 m/s relative to the mid-plane of the fluid. For the electro-osmotic (EOF) model, flow is induced by imposing an electric field of 0.05 V/nm \cite{a03,a08}. In the Poiseuille flow (PF) model flow is induced by a gravity field of $5\times10^{-5}$ kcal/(\AA$\cdot$gm) \cite{a18}. These dynamic models all start from the 20 ns output of the SM model and are carried out for an additional 20 ns when fluid velocity and charge distributions are converged. Data are output every 10 time steps in the last 10 ns of each simulation. The simulations were run in the Shenzhen National Computer Center of China.

The position of the slipping plane is determined most accurately with the CF model and this model determines the slipping plane for all the simulations reported here. The linear change in the velocity profile in the central portion of the fluid layer is fit to a straight line, which is then projected toward the solid surfaces. The distance from the surfaces at which the regression line velocity equals that of the solid surfaces defines the slipping plane.

The different flow mechanisms investigate the suggested origins of neutral surface zeta potentials. If the zeta potential arises from a surface charge in the solution immediately adjacent to the solid, the static model might predict the measured zeta potentials best. If the zeta potential arises from the different mobilities of the different ions in the bound water layer, one or several of the flow models should predict the best zeta potentials. 

\textit{Modeling results.} Figures 1 through 4 give the results. Each figure shows the averages of various parameters in thin surface-parallel planes as a function of distance from the solid surface. 
 
Figure 1 shows the results for a neutral monovalent surface submerged in 1 M NaCl solution under static and three dynamic flow conditions. The static and shear models (SM and CF) are similar within the solid and in the fluid to the slipping plane, but then the electrostatic potentials diverge slightly (Fig. 1a). The EOF model has distinctly lower electrostatic potential, and the PF model higher electrostatic potential in this interval. The water molecules are tightly bound to the numerical solid; the density of oxygen in water near the solid surface is $\sim$4.5 times that in the water at the mid-plane.  The concentration of ions is very different for the various flow systems (bottom panels).  Only the CF and SM models are similar. Based on these observations we conclude that: (1) the CF flow model can provide a reasonable determination of the slipping plane, and (2) the static model is the most realistic model for determining the zeta potential.

\begin{figure}
	\centering
		\includegraphics[width=0.50\textwidth]{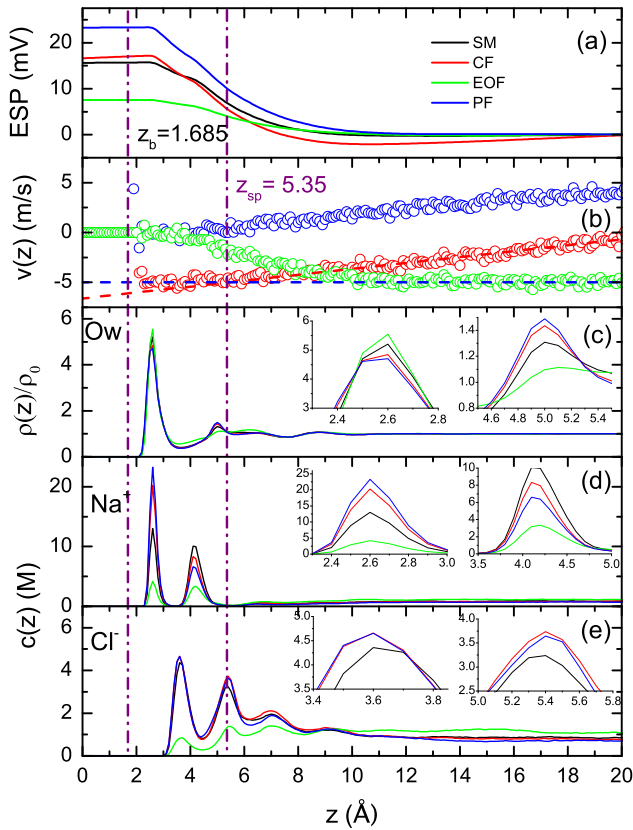}
	\caption{Comparison of calculated electrostatic, density, and concentration profiles for four flow systems across a neutral monovalent ($\mid e\mid=1$) surface submerged in 1 m NaCl electrolyte solution: Static model (SM-black), shear or Couette flow (CF-red), electro-osmotic flow (EOF-green) and Poiseuille flow (PF-blue). The vertical purple dot-dashed line is the slipping plane defined by the CF model; the dot-dashed line closer to the surface at z$_{b}$=1.685 \AA indicates top of the first row of atoms in the solid. The average concentration of Na$^{+}$ and Cl$^{-}$ ions on surface-parallel planes for the flow models are indicated by line color as indicated in panel (a). The inserts in (d) and (c) magnify the profiles across the ion peaks closest to the surface.}
	\label{fig:fig_1}
\end{figure}

Figure 2 compares the zeta potentials calculated using our procedure to those calculated using the H-S equation. In the top panels the fluid is 1M chloride electrolyte, and in the bottom panels the fluid is iodide electrolyte. The zeta potential calculated using the H-S methods are shown in the right panels; those calculated by our methods are on the left. We confirm Huang’s zeta potential of $-$38 mV for a zero-charge surface in NaI solution. The three data points that lie at $\mid e\mid =1.5$ represent experiments with an Al$_{2}$O$_{3}$ solid \cite{a21, a22}. Our method predicts zeta potentials within a few millivolts of those measured, and almost all the measured values fall within the prediction error bars, whereas the Helmholtz and Smoluchowski method predicts zeta potentials which are all more than 10 millivolts from those measured, and all fall well outside the prediction error bars. The curves in all the panels slope upward indicating that neutral surfaces with higher partial charge have higher zeta potentials. In chloride solutions (top panels), the Na+ and K+ cations only weakly interact with the $\mid e\mid=0$ surface (e.g., the zeta potential is near zero), but in iodide solutions (bottom panels) these same surfaces and cations have strongly negative zeta potentials. 

\begin{figure}
	\centering
		\includegraphics[width=0.50\textwidth]{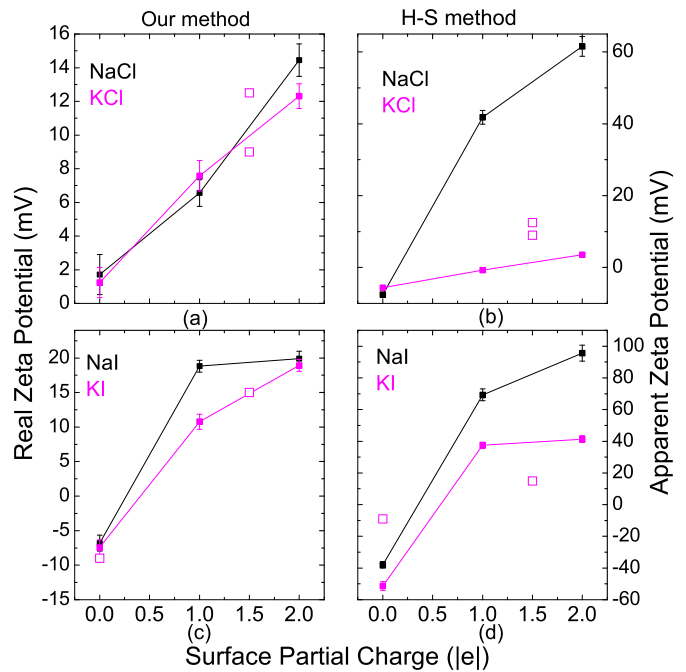}
	\caption{The calculated zeta potential for neutral surfaces with partial charges of $\mid e\mid $ = 0, 1, and 2 in contact with 1 M NaCl (top)  and 1 M NaI (bottom) are indicated by data points with error bars connected by lines. The hollow squares at $\mid e\mid=1.5$ represent the experimental measurements on Al$_{2}$O$_{3}$ \cite{a21, a22} ($\mid e\mid=1.5$), the hollow square at $\mid e\mid=0$ is a liposome bio-surface \cite{a09}.}
	\label{fig:fig_2}
\end{figure}

Figures 3 and 4 show that the ionic structure of the hydration layer is controlled mainly by the surface partial charge. Higher partial charge surfaces bind water more tightly and increases the water density. The $\rho_{Ow}$ peak (black curve) for the bottom ($\mid e\mid =2$) pair of panels indicates a relative density of oxygen in water near the solid surfaces of 5.5 to 6, whereas the density is $\sim$2.5 in the top pair of panels ($\mid e\mid =0$). For $\mid e\mid =0$ surfaces (top panels) there is almost no offset between the red hydrogen water molecule relative density peak and the black relative oxygen water density peak, but  for the monovalent and divalent solids (lower two panels) the red hydrogen density peak is closer to the surface. All the peaks are broadest for the zero-charge ($\mid e\mid =0$) surface.

The ion distribution strongly depends on whether the solution anion is chlorine or iodine. For the $\mid e\mid =0$ solid surfaces in the top panels of Figures 3 and 4 there are no  Na$^{+}$ or Cl$^{-}$ peaks near the surface for chloride solutions (left top panel), but there is a small Na$^{+}$ peak and a very large I$^{-}$ peak for the iodide solutions (top right panel). This greatly affects the zeta potential and is the reason that the zeta potential for $\mid e\mid =0$ surfaces is $\sim$0 in NaCl solutions but $\sim$$-$7.8 mV for NaI solutions.

\begin{figure}
	\centering
		\includegraphics[width=0.50\textwidth]{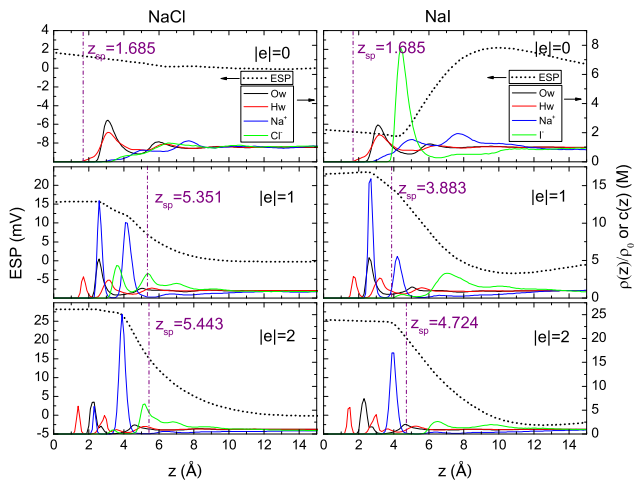}
	\caption{Simulated average relative density profiles, $\rho$(z)/$\rho_{bulk}$, of water oxygen (the black solid line), water hydrogen (red solid line), and average concentration profiles for positive (blue solid line) and negative (green solid line) ions, plotted as a function of distance z from the surface.  The centers of the first line of solid atoms is at $z=0$. The top pair of panels, $\mid e\mid=0$, is for a charge-neutral solid composed of zero partial charge atoms, the middle pair of panels for monovalent atoms, and the bottom pair  for divalent atoms. The left column of panels shows profiles for 1M solution of NaCl solutions, and  the right  column shows profiles for 1 M NaI solutions. The purple dash-dot line is the slipping plane determined by the CF models. }
	\label{fig:fig_3}
\end{figure}
 
The large size of the I$^{-}$ ion compared to the Cl$^{-}$ ion is the cause of these differences. The charge density of a large ion is smaller and it is more easily separated from water molecules [31].  When electrostatic forces are small or not present ($\mid e\mid =0$ and 1 surfaces in top panels) this size effect is most noticible. In contrast, the K$^{+}$ peaks are very similar to the Na$^{+}$ peaks because the ion sizes are very similar, which allows us to compare Huang \textit{et al.}’s predictions for the liposome bio-surface in 1M NaI to measurements in 1M KI solutions \cite{a03}. For the divalent neutral surface (bottom panels in Figs. 3 and 4) the water hydration layer is so tightly bound that there is only one Na$^{+}$ peak, and the ion size makes less difference. The I$^{-}$ ion size still increases the zeta potential, however, as shown in Figure 2. 

\begin{figure}
	\centering
		\includegraphics[width=0.50\textwidth]{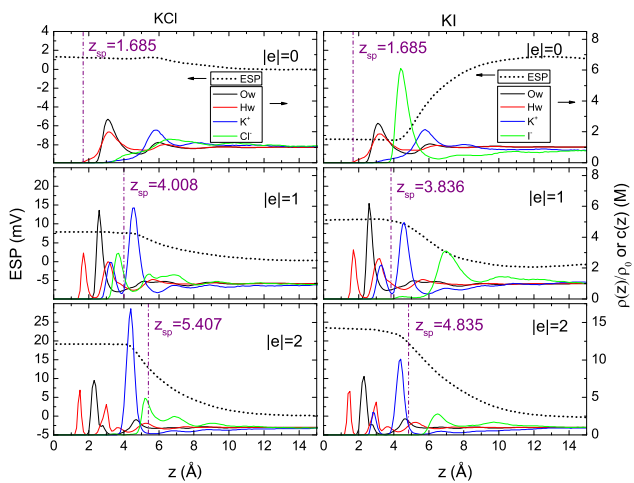}
	\caption{Same as Fig. 3 but for iodide rather than chloride solutions. See Figure 3 caption for discussion of curves.}
	\label{fig:fig_4}
\end{figure}

Finally it should be noted that the slipping plane is substantially further from the surface for the higher partial charge surfaces. The slipping plane is closer to the solid surface for the iodide compared to the chloride solutions (left vs right panels in Figs. 3 and 4). Whether the cation is Na$^{+}$ or K$^{+}$ makes little difference, however. These shifts in the slipping plane do not affect the zeta potential as much as one might expect because the electrostatic potential profile is fairly flat close to the surface.

\textit{Discussion.} The method of integration used here differs from that used by earlier workers in that it averages the charge distribution in each incremental layer prior to integrating the averaged charge layer to obtain the electrostatic potential rather than integrating along a number of lines perpendicular to the surface and then averaging their potentials. Our method appears to be better because it shows none of the non-physical oscillations that plague the other methods.

The accuracy of the electrostatic integration depends on the thickness of the incremental charge sheets, which we take as 0.01 nm. The sheets need to be thick enough to capture a representative number of ions, but thin enough to provide smooth integration.  The electrostatic potential determined with a sheet thicknesses of 0.01 nm is similar to that obtained with 0.001 nm and both have better resolution than integrating with 0.025 nm sheets (results not shown here).  On this basis we believe our choice of charge layer thickness is appropriate.
 
The position of slipping plane was determined using shearing velocities of $\pm$5 m/s (4.64$\pm$1.58 \AA) in the CF model. Shearing velocities of $\pm$1 m/s (5.25$\pm$1.13 \AA) and $\pm$10 m/s  (5.07$\pm$0.46 \AA) indicate similar positions of slipping plan and show overlapping errors which are larger for the smaller shearing velocities where the thermal scatter is proportionately greater. The uncertainties in the slipping plane position at $\pm$5m/s are acceptably small for the conclusions reached in this paper.

Ion binding to the surface is the main reason that neutral surfaces have non-zero zeta potential. Ion penetration of the hydration layer makes a secondary contribution. Small ions such as Na$^{+}$ penetrate the hydration layer more than larger ions such as K$^{+}$ (Figures 3 and 4).

Finally the force field used in the simulations can be important and selecting the right force field for when applications are made of our method may be challenging.  We use the force field of Huang \textit{et al.} \cite{a03}.  The agreement between our calculated and measured zeta potentials for neutral surfaces (Fig. 2) suggest this force file is appropriate, but the agreement could a coincidence.  Further work will be needed to completely eliminate this possibility.

\textit{Conclusion.} We demonstrate how MD methods can be used to calculate zeta potentials of neutral surfaces comprised of zero-charge, monovalent, and divalent ions that are close enough to measured values that their distribution of atoms can be considered potentially instructive.  Higher surface partial charge results in higher zeta potentials (Fig. 2).  Cation size has a strong control ( left vs right panels in Figs. 3 and 4). The zeta potential of neutral surfaces is influenced by the structure of the hydration layer, the surface partial charge, and the size of aqueous ions and this is perhaps best captured by MD methods.  If confirmed by further work, we believe the methods described here could significantly increase the utility of MD modeling in understanding surface chemical phenomena.

\begin{acknowledgments}
We acknowledge the helpful advice provided by Professor Athanassios Z. Panagiotopoulos of Princeton University. This publication was funded by Award No. KUS-C1-018-02 from the King Abdullah University of Science and Technology (KAUST). We are grateful for their support.
\end{acknowledgments}

\bibliography{PRLms}

\begin{thebibliography}{29}%
\makeatletter
\providecommand \@ifxundefined [1]{%
 \@ifx{#1\undefined}
}%
\providecommand \@ifnum [1]{%
 \ifnum #1\expandafter \@firstoftwo
 \else \expandafter \@secondoftwo
 \fi
}%
\providecommand \@ifx [1]{%
 \ifx #1\expandafter \@firstoftwo
 \else \expandafter \@secondoftwo
 \fi
}%
\providecommand \natexlab [1]{#1}%
\providecommand \enquote  [1]{``#1''}%
\providecommand \bibnamefont  [1]{#1}%
\providecommand \bibfnamefont [1]{#1}%
\providecommand \citenamefont [1]{#1}%
\providecommand \href@noop [0]{\@secondoftwo}%
\providecommand \href [0]{\begingroup \@sanitize@url \@href}%
\providecommand \@href[1]{\@@startlink{#1}\@@href}%
\providecommand \@@href[1]{\endgroup#1\@@endlink}%
\providecommand \@sanitize@url [0]{\catcode `\\12\catcode `\$12\catcode
  `\&12\catcode `\#12\catcode `\^12\catcode `\_12\catcode `\%12\relax}%
\providecommand \@@startlink[1]{}%
\providecommand \@@endlink[0]{}%
\providecommand \url  [0]{\begingroup\@sanitize@url \@url }%
\providecommand \@url [1]{\endgroup\@href {#1}{\urlprefix }}%
\providecommand \urlprefix  [0]{URL }%
\providecommand \Eprint [0]{\href }%
\providecommand \doibase [0]{http://dx.doi.org/}%
\providecommand \selectlanguage [0]{\@gobble}%
\providecommand \bibinfo  [0]{\@secondoftwo}%
\providecommand \bibfield  [0]{\@secondoftwo}%
\providecommand \translation [1]{[#1]}%
\providecommand \BibitemOpen [0]{}%
\providecommand \bibitemStop [0]{}%
\providecommand \bibitemNoStop [0]{.\EOS\space}%
\providecommand \EOS [0]{\spacefactor3000\relax}%
\providecommand \BibitemShut  [1]{\csname bibitem#1\endcsname}%
\let\auto@bib@innerbib\@empty
\bibitem [{\citenamefont {Huang}\ \emph {et~al.}(2007)\citenamefont {Huang},
  \citenamefont {Cottin-Bizonne}, \citenamefont {Ybert},\ and\ \citenamefont
  {Bocquet}}]{a03}%
  \BibitemOpen
  \bibfield  {author} {\bibinfo {author} {\bibfnamefont {D.~M.}\ \bibnamefont
  {Huang}}, \bibinfo {author} {\bibfnamefont {C.}~\bibnamefont
  {Cottin-Bizonne}}, \bibinfo {author} {\bibfnamefont {C.}~\bibnamefont
  {Ybert}}, \ and\ \bibinfo {author} {\bibfnamefont {L.}~\bibnamefont
  {Bocquet}},\ }\href@noop {} {\bibfield  {journal} {\bibinfo  {journal} {Phys
  Rev Lett}\ }\textbf {\bibinfo {volume} {98}},\ \bibinfo {pages} {177801}
  (\bibinfo {year} {2007})}\BibitemShut {NoStop}%
\bibitem [{\citenamefont {Joly}\ \emph {et~al.}(2004)\citenamefont {Joly},
  \citenamefont {Ybert}, \citenamefont {Trizac},\ and\ \citenamefont
  {Bocquet}}]{a04}%
  \BibitemOpen
  \bibfield  {author} {\bibinfo {author} {\bibfnamefont {L.}~\bibnamefont
  {Joly}}, \bibinfo {author} {\bibfnamefont {C.}~\bibnamefont {Ybert}},
  \bibinfo {author} {\bibfnamefont {E.}~\bibnamefont {Trizac}}, \ and\ \bibinfo
  {author} {\bibfnamefont {L.}~\bibnamefont {Bocquet}},\ }\href@noop {}
  {\bibfield  {journal} {\bibinfo  {journal} {Phys. Rev. Lett.}\ }\textbf
  {\bibinfo {volume} {93}},\ \bibinfo {pages} {257805/1} (\bibinfo {year}
  {2004})}\BibitemShut {NoStop}%
\bibitem [{\citenamefont {Kirby}\ and\ \citenamefont
  {Hasselbrink}(2004)}]{a05}%
  \BibitemOpen
  \bibfield  {author} {\bibinfo {author} {\bibfnamefont {B.~J.}\ \bibnamefont
  {Kirby}}\ and\ \bibinfo {author} {\bibfnamefont {E.~F.}\ \bibnamefont
  {Hasselbrink}},\ }\href@noop {} {\bibfield  {journal} {\bibinfo  {journal}
  {Electrophoresis}\ }\textbf {\bibinfo {volume} {25}},\ \bibinfo {pages} {187}
  (\bibinfo {year} {2004})}\BibitemShut {NoStop}%
\bibitem [{\citenamefont {Lyklema}(2011)}]{a06}%
  \BibitemOpen
  \bibfield  {author} {\bibinfo {author} {\bibfnamefont {J.}~\bibnamefont
  {Lyklema}},\ }\href@noop {} {\bibfield  {journal} {\bibinfo  {journal}
  {Colloids and Surfaces A}\ }\textbf {\bibinfo {volume} {376}},\ \bibinfo
  {pages} {2} (\bibinfo {year} {2011})}\BibitemShut {NoStop}%
\bibitem [{\citenamefont {Booth}(1948)}]{a01}%
  \BibitemOpen
  \bibfield  {author} {\bibinfo {author} {\bibfnamefont {F.}~\bibnamefont
  {Booth}},\ }\href@noop {} {\bibfield  {journal} {\bibinfo  {journal}
  {Nature}\ }\textbf {\bibinfo {volume} {161}},\ \bibinfo {pages} {83}
  (\bibinfo {year} {1948})}\BibitemShut {NoStop}%
\bibitem [{\citenamefont {Joly}\ \emph {et~al.}(2006)\citenamefont {Joly},
  \citenamefont {Ybert}, \citenamefont {Trizac},\ and\ \citenamefont
  {Bocquet}}]{a07}%
  \BibitemOpen
  \bibfield  {author} {\bibinfo {author} {\bibfnamefont {L.}~\bibnamefont
  {Joly}}, \bibinfo {author} {\bibfnamefont {C.}~\bibnamefont {Ybert}},
  \bibinfo {author} {\bibfnamefont {E.}~\bibnamefont {Trizac}}, \ and\ \bibinfo
  {author} {\bibfnamefont {L.}~\bibnamefont {Bocquet}},\ }\href@noop {}
  {\bibfield  {journal} {\bibinfo  {journal} {J. Chem. Phys.}\ }\textbf
  {\bibinfo {volume} {125}},\ \bibinfo {pages} {204716/1} (\bibinfo {year}
  {2006})}\BibitemShut {NoStop}%
\bibitem [{\citenamefont {Huang}\ \emph {et~al.}(2008)\citenamefont {Huang},
  \citenamefont {Cottin-Bizonne}, \citenamefont {Ybert},\ and\ \citenamefont
  {Bocquet}}]{a08}%
  \BibitemOpen
  \bibfield  {author} {\bibinfo {author} {\bibfnamefont {D.~M.}\ \bibnamefont
  {Huang}}, \bibinfo {author} {\bibfnamefont {C.}~\bibnamefont
  {Cottin-Bizonne}}, \bibinfo {author} {\bibfnamefont {C.}~\bibnamefont
  {Ybert}}, \ and\ \bibinfo {author} {\bibfnamefont {L.}~\bibnamefont
  {Bocquet}},\ }\href@noop {} {\bibfield  {journal} {\bibinfo  {journal}
  {Langmuir}\ }\textbf {\bibinfo {volume} {24}},\ \bibinfo {pages} {1442}
  (\bibinfo {year} {2008})}\BibitemShut {NoStop}%
\bibitem [{\citenamefont {Jelinek}\ \emph {et~al.}(2011)\citenamefont
  {Jelinek}, \citenamefont {Felicelli}, \citenamefont {Mlakar},\ and\
  \citenamefont {Peters}}]{ajj}%
  \BibitemOpen
  \bibfield  {author} {\bibinfo {author} {\bibfnamefont {B.}~\bibnamefont
  {Jelinek}}, \bibinfo {author} {\bibfnamefont {S.~D.}\ \bibnamefont
  {Felicelli}}, \bibinfo {author} {\bibfnamefont {P.~F.}\ \bibnamefont
  {Mlakar}}, \ and\ \bibinfo {author} {\bibfnamefont {J.~F.}\ \bibnamefont
  {Peters}},\ }\href@noop {} {\bibfield  {journal} {\bibinfo  {journal}
  {International Journal of Theoretical and Applied Multiscale Mechanics}\
  }\textbf {\bibinfo {volume} {2}} (\bibinfo {year} {2011})}\BibitemShut
  {NoStop}%
\bibitem [{\citenamefont {Petrache}\ \emph {et~al.}(2006)\citenamefont
  {Petrache}, \citenamefont {Zemb}, \citenamefont {Belloni},\ and\
  \citenamefont {Parsegian}}]{a09}%
  \BibitemOpen
  \bibfield  {author} {\bibinfo {author} {\bibfnamefont {H.~I.}\ \bibnamefont
  {Petrache}}, \bibinfo {author} {\bibfnamefont {T.}~\bibnamefont {Zemb}},
  \bibinfo {author} {\bibfnamefont {L.}~\bibnamefont {Belloni}}, \ and\
  \bibinfo {author} {\bibfnamefont {V.~A.}\ \bibnamefont {Parsegian}},\
  }\href@noop {} {\bibfield  {journal} {\bibinfo  {journal} {Proceedings of the
  National Academy of Sciences}\ }\textbf {\bibinfo {volume} {103}},\ \bibinfo
  {pages} {7982} (\bibinfo {year} {2006})}\BibitemShut {NoStop}%
\bibitem [{\citenamefont {Qiao}\ and\ \citenamefont {Aluru}(2003)}]{a10}%
  \BibitemOpen
  \bibfield  {author} {\bibinfo {author} {\bibfnamefont {R.}~\bibnamefont
  {Qiao}}\ and\ \bibinfo {author} {\bibfnamefont {N.~R.}\ \bibnamefont
  {Aluru}},\ }\href@noop {} {\bibfield  {journal} {\bibinfo  {journal} {J Chem
  Phys}\ }\textbf {\bibinfo {volume} {118}},\ \bibinfo {pages} {4692} (\bibinfo
  {year} {2003})}\BibitemShut {NoStop}%
\bibitem [{\citenamefont {Qiao}\ and\ \citenamefont {Aluru}(2004)}]{a11}%
  \BibitemOpen
  \bibfield  {author} {\bibinfo {author} {\bibfnamefont {R.}~\bibnamefont
  {Qiao}}\ and\ \bibinfo {author} {\bibfnamefont {N.~R.}\ \bibnamefont
  {Aluru}},\ }\href@noop {} {\bibfield  {journal} {\bibinfo  {journal} {Phys
  Rev Lett}\ }\textbf {\bibinfo {volume} {92}},\ \bibinfo {pages} {198301}
  (\bibinfo {year} {2004})}\BibitemShut {NoStop}%
\bibitem [{\citenamefont {Khan}\ \emph {et~al.}(2005)\citenamefont {Khan},
  \citenamefont {Petris},\ and\ \citenamefont {Chan}}]{a12}%
  \BibitemOpen
  \bibfield  {author} {\bibinfo {author} {\bibfnamefont {M.~O.}\ \bibnamefont
  {Khan}}, \bibinfo {author} {\bibfnamefont {S.}~\bibnamefont {Petris}}, \ and\
  \bibinfo {author} {\bibfnamefont {D.~Y.~C.}\ \bibnamefont {Chan}},\
  }\href@noop {} {\bibfield  {journal} {\bibinfo  {journal} {J Chem Phys}\
  }\textbf {\bibinfo {volume} {122}},\ \bibinfo {pages} {104705} (\bibinfo
  {year} {2005})}\BibitemShut {NoStop}%
\bibitem [{\citenamefont {Diehl}\ and\ \citenamefont {Levin}(2006)}]{a13}%
  \BibitemOpen
  \bibfield  {author} {\bibinfo {author} {\bibfnamefont {A.}~\bibnamefont
  {Diehl}}\ and\ \bibinfo {author} {\bibfnamefont {Y.}~\bibnamefont {Levin}},\
  }\href@noop {} {\bibfield  {journal} {\bibinfo  {journal} {The Journal of
  Chemical Physics}\ }\textbf {\bibinfo {volume} {125}},\ \bibinfo {pages}
  {054902} (\bibinfo {year} {2006})}\BibitemShut {NoStop}%
\bibitem [{\citenamefont {Lorenz}\ \emph {et~al.}(2008)\citenamefont {Lorenz},
  \citenamefont {Crozier}, \citenamefont {Anderson},\ and\ \citenamefont
  {Travesset}}]{a18}%
  \BibitemOpen
  \bibfield  {author} {\bibinfo {author} {\bibfnamefont {C.~D.}\ \bibnamefont
  {Lorenz}}, \bibinfo {author} {\bibfnamefont {P.~S.}\ \bibnamefont {Crozier}},
  \bibinfo {author} {\bibfnamefont {J.~A.}\ \bibnamefont {Anderson}}, \ and\
  \bibinfo {author} {\bibfnamefont {A.}~\bibnamefont {Travesset}},\ }\href@noop
  {} {\bibfield  {journal} {\bibinfo  {journal} {J. Phys. Chem. C}\ }\textbf
  {\bibinfo {volume} {112}},\ \bibinfo {pages} {10222} (\bibinfo {year}
  {2008})}\BibitemShut {NoStop}%
\bibitem [{\citenamefont {Spohr}(1997)}]{a14}%
  \BibitemOpen
  \bibfield  {author} {\bibinfo {author} {\bibfnamefont {E.}~\bibnamefont
  {Spohr}},\ }\href@noop {} {\bibfield  {journal} {\bibinfo  {journal} {J Chem
  Phys}\ }\textbf {\bibinfo {volume} {107}},\ \bibinfo {pages} {6342} (\bibinfo
  {year} {1997})}\BibitemShut {NoStop}%
\bibitem [{\citenamefont {Spohr}(1999)}]{a15}%
  \BibitemOpen
  \bibfield  {author} {\bibinfo {author} {\bibfnamefont {E.}~\bibnamefont
  {Spohr}},\ }\href@noop {} {\bibfield  {journal} {\bibinfo  {journal}
  {Electrochimica Acta}\ }\textbf {\bibinfo {volume} {44}},\ \bibinfo {pages}
  {1697} (\bibinfo {year} {1999})}\BibitemShut {NoStop}%
\bibitem [{\citenamefont {Spagnoli}\ \emph {et~al.}(2006)\citenamefont
  {Spagnoli}, \citenamefont {Cooke}, \citenamefont {Kerisit},\ and\
  \citenamefont {Parker}}]{a19}%
  \BibitemOpen
  \bibfield  {author} {\bibinfo {author} {\bibfnamefont {D.}~\bibnamefont
  {Spagnoli}}, \bibinfo {author} {\bibfnamefont {D.~J.}\ \bibnamefont {Cooke}},
  \bibinfo {author} {\bibfnamefont {S.}~\bibnamefont {Kerisit}}, \ and\
  \bibinfo {author} {\bibfnamefont {S.~C.}\ \bibnamefont {Parker}},\
  }\href@noop {} {\bibfield  {journal} {\bibinfo  {journal} {Journal of
  Materials Chemistry}\ }\textbf {\bibinfo {volume} {16}},\ \bibinfo {pages}
  {1997} (\bibinfo {year} {2006})}\BibitemShut {NoStop}%
\bibitem [{\citenamefont {Bagotskii}(2006)}]{a16}%
  \BibitemOpen
  \bibfield  {author} {\bibinfo {author} {\bibfnamefont {V.~S.}\ \bibnamefont
  {Bagotskii}},\ }\href@noop {} {\emph {\bibinfo {title} {Fundamentals of
  electrochemistry}}}\ (\bibinfo  {publisher} {Wiley-Interscience},\ \bibinfo
  {address} {Hoboken, N.J.},\ \bibinfo {year} {2006})\BibitemShut {NoStop}%
\bibitem [{\citenamefont {Kirby}(2010)}]{a17}%
  \BibitemOpen
  \bibfield  {author} {\bibinfo {author} {\bibfnamefont {B.~J.}\ \bibnamefont
  {Kirby}},\ }\href@noop {} {\emph {\bibinfo {title} {Micro- and nanoscale
  fluid mechanics : transport in microfluidic devices}}}\ (\bibinfo
  {publisher} {Cambridge University Press},\ \bibinfo {address} {New York},\
  \bibinfo {year} {2010})\BibitemShut {NoStop}%
\bibitem [{\citenamefont {Kosmulski}\ and\ \citenamefont
  {Rosenholm}(1996)}]{a20}%
  \BibitemOpen
  \bibfield  {author} {\bibinfo {author} {\bibfnamefont {M.}~\bibnamefont
  {Kosmulski}}\ and\ \bibinfo {author} {\bibfnamefont {J.~B.}\ \bibnamefont
  {Rosenholm}},\ }\href@noop {} {\bibfield  {journal} {\bibinfo  {journal}
  {Journal of Physical Chemistry}\ }\textbf {\bibinfo {volume} {100}},\
  \bibinfo {pages} {11681} (\bibinfo {year} {1996})}\BibitemShut {NoStop}%
\bibitem [{\citenamefont {Johnson}\ \emph {et~al.}(1999)\citenamefont
  {Johnson}, \citenamefont {Scales},\ and\ \citenamefont {Healy}}]{a21}%
  \BibitemOpen
  \bibfield  {author} {\bibinfo {author} {\bibfnamefont {S.~B.}\ \bibnamefont
  {Johnson}}, \bibinfo {author} {\bibfnamefont {P.~J.}\ \bibnamefont {Scales}},
  \ and\ \bibinfo {author} {\bibfnamefont {T.~W.}\ \bibnamefont {Healy}},\
  }\href@noop {} {\bibfield  {journal} {\bibinfo  {journal} {Langmuir}\
  }\textbf {\bibinfo {volume} {15}},\ \bibinfo {pages} {2836} (\bibinfo {year}
  {1999})}\BibitemShut {NoStop}%
\bibitem [{\citenamefont {Dukhin}\ \emph {et~al.}(2005)\citenamefont {Dukhin},
  \citenamefont {Dukhin},\ and\ \citenamefont {Goetz}}]{a22}%
  \BibitemOpen
  \bibfield  {author} {\bibinfo {author} {\bibfnamefont {A.}~\bibnamefont
  {Dukhin}}, \bibinfo {author} {\bibfnamefont {S.}~\bibnamefont {Dukhin}}, \
  and\ \bibinfo {author} {\bibfnamefont {P.}~\bibnamefont {Goetz}},\
  }\href@noop {} {\bibfield  {journal} {\bibinfo  {journal} {Langmuir}\
  }\textbf {\bibinfo {volume} {21}},\ \bibinfo {pages} {9990} (\bibinfo {year}
  {2005})}\BibitemShut {NoStop}%
\bibitem [{\citenamefont {Heberling}\ \emph {et~al.}(2011)\citenamefont
  {Heberling}, \citenamefont {Trainor}, \citenamefont {Lützenkirchen},
  \citenamefont {Eng}, \citenamefont {Denecke},\ and\ \citenamefont
  {Bosbach}}]{a23}%
  \BibitemOpen
  \bibfield  {author} {\bibinfo {author} {\bibfnamefont {F.}~\bibnamefont
  {Heberling}}, \bibinfo {author} {\bibfnamefont {T.~P.}\ \bibnamefont
  {Trainor}}, \bibinfo {author} {\bibfnamefont {J.}~\bibnamefont
  {Lützenkirchen}}, \bibinfo {author} {\bibfnamefont {P.}~\bibnamefont {Eng}},
  \bibinfo {author} {\bibfnamefont {M.~A.}\ \bibnamefont {Denecke}}, \ and\
  \bibinfo {author} {\bibfnamefont {D.}~\bibnamefont {Bosbach}},\ }\href@noop
  {} {\bibfield  {journal} {\bibinfo  {journal} {Journal of Colloid and
  Interface Science}\ }\textbf {\bibinfo {volume} {354}},\ \bibinfo {pages}
  {843} (\bibinfo {year} {2011})}\BibitemShut {NoStop}%
\bibitem [{\citenamefont {Joseph}\ and\ \citenamefont {Aluru}(2006)}]{a24}%
  \BibitemOpen
  \bibfield  {author} {\bibinfo {author} {\bibfnamefont {S.}~\bibnamefont
  {Joseph}}\ and\ \bibinfo {author} {\bibfnamefont {N.~R.}\ \bibnamefont
  {Aluru}},\ }\href@noop {} {\bibfield  {journal} {\bibinfo  {journal}
  {Langmuir}\ }\textbf {\bibinfo {volume} {22}},\ \bibinfo {pages} {9041}
  (\bibinfo {year} {2006})}\BibitemShut {NoStop}%
\bibitem [{\citenamefont {Koneshan}\ \emph {et~al.}(1998)\citenamefont
  {Koneshan}, \citenamefont {Rasaiah}, \citenamefont {Lynden-Bell},\ and\
  \citenamefont {Lee}}]{a25}%
  \BibitemOpen
  \bibfield  {author} {\bibinfo {author} {\bibfnamefont {S.}~\bibnamefont
  {Koneshan}}, \bibinfo {author} {\bibfnamefont {J.~C.}\ \bibnamefont
  {Rasaiah}}, \bibinfo {author} {\bibfnamefont {R.~M.}\ \bibnamefont
  {Lynden-Bell}}, \ and\ \bibinfo {author} {\bibfnamefont {S.~H.}\ \bibnamefont
  {Lee}},\ }\href@noop {} {\bibfield  {journal} {\bibinfo  {journal} {Journal
  of Physical Chemistry B}\ }\textbf {\bibinfo {volume} {102}},\ \bibinfo
  {pages} {4193} (\bibinfo {year} {1998})}\BibitemShut {NoStop}%
\bibitem [{\citenamefont {Kerisit}\ and\ \citenamefont {Liu}(2009)}]{a26}%
  \BibitemOpen
  \bibfield  {author} {\bibinfo {author} {\bibfnamefont {S.}~\bibnamefont
  {Kerisit}}\ and\ \bibinfo {author} {\bibfnamefont {C.}~\bibnamefont {Liu}},\
  }\href@noop {} {\bibfield  {journal} {\bibinfo  {journal} {Environmental
  Science and Technology}\ }\textbf {\bibinfo {volume} {43}},\ \bibinfo {pages}
  {777} (\bibinfo {year} {2009})}\BibitemShut {NoStop}%
\bibitem [{\citenamefont {Bourg}\ and\ \citenamefont {Sposito}(2011)}]{a27}%
  \BibitemOpen
  \bibfield  {author} {\bibinfo {author} {\bibfnamefont {I.~C.}\ \bibnamefont
  {Bourg}}\ and\ \bibinfo {author} {\bibfnamefont {G.}~\bibnamefont
  {Sposito}},\ }\href@noop {} {\bibfield  {journal} {\bibinfo  {journal} {J
  Colloid Interface Sci}\ }\textbf {\bibinfo {volume} {360}},\ \bibinfo {pages}
  {701} (\bibinfo {year} {2011})}\BibitemShut {NoStop}%
\bibitem [{\citenamefont {Yeh}\ and\ \citenamefont {Berkowitz}(1999)}]{a28}%
  \BibitemOpen
  \bibfield  {author} {\bibinfo {author} {\bibfnamefont {I.-C.}\ \bibnamefont
  {Yeh}}\ and\ \bibinfo {author} {\bibfnamefont {M.~L.}\ \bibnamefont
  {Berkowitz}},\ }\href@noop {} {\bibfield  {journal} {\bibinfo  {journal} {The
  Journal of Chemical Physics}\ }\textbf {\bibinfo {volume} {111}},\ \bibinfo
  {pages} {3155} (\bibinfo {year} {1999})}\BibitemShut {NoStop}%
\bibitem [{\citenamefont {Plimpton}(1995)}]{a29}%
  \BibitemOpen
  \bibfield  {author} {\bibinfo {author} {\bibfnamefont {S.}~\bibnamefont
  {Plimpton}},\ }\href@noop {} {\bibfield  {journal} {\bibinfo  {journal}
  {Journal of Computational Physics}\ }\textbf {\bibinfo {volume} {117}},\
  \bibinfo {pages} {1} (\bibinfo {year} {1995})}\BibitemShut {NoStop}%
\end{thebibliography}%

\end{document}